\documentclass[12pt]{revtex4}
\usepackage[T1]{fontenc}
\usepackage{graphicx}

\begin{document}

\title[In-plane optical anisotropy]{In-plane optical anisotropy due to conduction band
electron wavefunctions}

\author{J {\L}usakowski$^1$, M Sakowicz$^1$, K J Friedland$^2$, R Hey$^2$ and K~Ploog$^2$}

\address{$^1$ Institute of Experimental Physics, Warsaw University,
Ho\.za 69, 00-681 Warsaw, Poland}

\address{$^2$ Paul-Drude-Institut f\"{u}r Festk\"{o}rperelektronik,
Hausvogteiplatz 5-7, 10117 Berlin, Germany}

\begin{abstract}
Photoluminescence measurements were carried out on Be $\delta$-doped
GaAs/Al$_{0.33}$Ga$_{0.67}$As heterostructure at 1.6 K in magnetic
fields ($B$) up to 5 T. Luminescence originating from recombination
of a two-dimensional electron gas (2DEG) and photo excited holes
localized on Be acceptors was analyzed. The degree of circular
polarization ($\gamma_C$) of the luminescence from fully occupied
Landau levels was determined as a function of $B$ and the 2DEG
concentration, $n_s$. At $B$ constant, $\gamma_C$ decreased with the
increase of $n_s$. Two mechanisms of the $\gamma_C(n_s)$ dependence
are discussed: a) the Stark effect on a photo excited hole bound to
Be acceptor and b) the in-plane anisotropy of the intensity of
optical transitions. A quantitative analysis shows that the
influence of the Stark effect on $\gamma_C$ is negligible in the
present experiment. We propose that the $\gamma_C(n_s)$ dependence
results from the $C_{2v}$ symmetry of conduction band electron
wavefunctions and we give qualitative arguments supporting this
interpretation.
\end{abstract}

\maketitle

\section{Introduction}

Let us consider a GaAs/AlGaAs heterostructure. The point symmetry
group of the constituent semiconductors is $T_d$, but the presence
of the interface reduces the symmetry of the heterostructure to
$C_{2v}$. Lowering the  symmetry has important consequences on
optical properties, leading to an anisotropy in the (001) plane.
This anisotropy is of the current interest and is studied both
experimentally and theoretically. The landmark experiments were
carried out by Jusserand {\em et al.} \cite{Jusserand1, Jusserand2},
Kwok {\em et al.} \cite{Kwok}, Krebs and Voisin \cite{Krebs}, and
Kudelski {\em et al.} \cite{Kudelski}. These measurements were
accompanied by theoretical considerations concentrating on the
anisotropic spin splitting of the conduction band \cite{Andrada,
Pfeffer}, the anisotropy of the Raman scattering \cite{Froltsov} and
the anisotropy due to the heavy hole - light hole mixing by the
interface potential \cite{Ivchenko2}. A comprehensive review of
experimental data and theoretical models developed to investigate
the anisotropy of heterostructures and quantum wells can be found in
\cite{Winkler}.

In the present paper we consider an influence of the $T_d
\rightarrow C_{2v}$ symmetry lowering  on the degree of circular
polarization of the luminescence originating from a single
GaAs/AlGaAs heterostructure. The investigated heterstructure was
grown in the [001] direction. A high mobility quasi two dimensional
electron gas (2DEG) is located at the interface due to Si donor
doping of the barrier. A diluted sheet of Be acceptors is introduced
into the GaAs layer at $z_0$ = 30 nm from the interface. The
photoluminescence analyzed originated from transitions between the
2DEG and photo excited holes bound to Be acceptors. Application of
the magnetic field ($B$) perpendicular to the (001) plane splits the
electron and hole levels which leads to a nonzero degree of circular
polarization at $B \neq 0$ and at sufficiently low temperatures.

\begin{figure} \begin{center}
\includegraphics*[scale=0.8,angle=0]{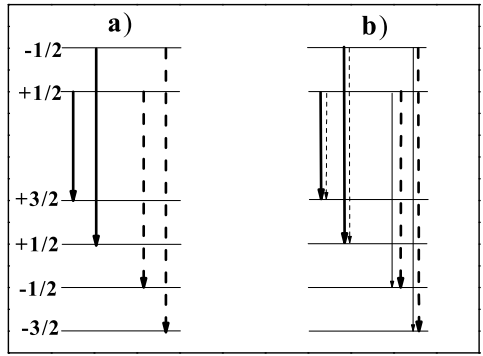}
\caption{a) Polarization selection rules for the radiative
recombination of a $\Gamma_6$ electron with a $\Gamma_8$ hole in the
$T_d$ symmetry. Solid arrows: $\sigma^-$, dashed arrows:
$\sigma^+$transitions. b) Lowering the symmetry of the system leads
to a relaxation of the selection rules and mixing of polarization of
transitions between given pair of levels.  Intensity of transitions
is mimicked schematically by lines thickness. Hole levels are not
 eigenstates of $J= 3/2$ orbital momentum operator but no
possible perturbation - related energy shifts are indicated.}
\label{JL1}
\end{center} \end{figure}

A near-band luminescence observed in 2D structures grown from zinc
blende materials is often interpreted as a $\Gamma_6 \rightarrow
\Gamma_8$ transition, where $\Gamma_6$ (conduction band) and
$\Gamma_8$ (valence band) are representations of the $T_d$ symmetry
group of the bulk crystal at zero wavevector $\mathbf{k}$. The
corresponding optical selection rules are shown in figure
\ref{JL1}a. In such a case, the degree of circular polarization of
the luminescence, $\gamma_C$, is determined only by the magnetic
field and the temperature ($T$): the magnetic field determines the
value of the Zeeman splitting of the electron and hole levels while
the temperature determines their occupation. The selection rules
shown in figure \ref{JL1}a are valid for the perfect $T_d$ symmetry
of the electron and hole states. A single heterojunction is a
structure of the $C_{2v}$ symmetry and the selection rules of figure
\ref{JL1}a are relaxed. The resulting scheme (derived further on) is
shown in figure \ref{JL1}b. Such a relaxation of the selection rules
leads to a decrease of the degree of circular polarization.

The origin of the present paper was an observation of a dependence
of $\gamma_C$ on the concentration of the 2DEG ($n_s$) which cannot
be explained on the basis of the selection rules shown in figure
\ref{JL1}a. To explain the effect observed, we discuss two
mechanisms which are related to the heterostructure electric field
and lead to selection rules shown in figure \ref{JL1}b. The first
one is the Stark effect on a photo excited hole bound to Be
acceptor. The second mechanism is an in-plane anisotropy of optical
transitions induced by a low symmetry of the heterostructure
potential.

The paper is organized as the following. To introduce the idea of
the present experiment, we start, in Section 2, with a description
of basic facts concerning polarization of the luminescence in
acceptor $\delta$-doped heterostructures. We follow, in Section 3
and Section 4, with the experimental part describing the sample
investigated, the experimental set-up and results. Next, in Section
5, we introduce a model of the Stark effect on the $\Gamma_8$ hole.
Application of the model to the experimental data leads to the
conclusion that the Stark effect cannot be responsible for the
$\gamma_C(n_s)$ dependence observed. Then, in Section 6, we propose
that it results from a low symmetry of the heterojunction
electrostatic potential. We argue that the anisotropy appears due to
conduction band electron wavefunctions, and not wavefunctions of
holes taking part in the optical transition. Within a
phenomenological model we derive appropriate selection rules shown
in figure \ref{JL1}b. Finally, we summarize and conclude the paper.

\section{Polarized luminescence from acceptor $\delta$-doped heterostructures}

Heterostructures similar to that investigated in the present work
were already studied by other groups \cite{KukushkinAdvances}. We
recall here some basic facts that are relevant to understand the
present experiments and results.

In acceptor $\delta$-doped heterostructures, acceptors are
introduced into the GaAs channel some tens of nanometers from the
GaAs/AlGaAs interface. A 2DEG is created in a heterostructure
quantum well due to doping of the AlGaAs barrier with shallow Si
donors. The photoluminescence (PL) is excited by a laser beam and
some of photo excited holes are captured by acceptors in the
$\delta$-layer. A typical luminescence spectrum shows a number of
lines corresponding to different transitions, like donor - acceptor
or exciton ones. Here we focus only on the radiative recombination
of 2D electrons with holes bound to acceptors in the $\delta$-layer.
Since this is the only optical transition analyzed in this paper,
the notion "hole" will always have the meaning of a hole bound to an
ionized acceptor in the $\delta$-layer.

If the magnetic field is applied perpendicular to the 2DEG plane,
the electron density of states is quantized into Landau levels (LLs)
and the degeneracy of the hole ground state is totally removed. In
such a case, the PL arising from the recombination of 2D electrons
with holes is polarized. The degree of circular polarization
 $\gamma_C= (I_{\sigma^{-}} -I_{\sigma^{+}})/(I_{\sigma^{-}} +I_{\sigma^{+}})$,
where $I_{\sigma^{+}}$ and $I_{\sigma^{-}}$ are intensities of the
$\sigma^+$ and $\sigma^-$ components of the luminescence,
respectively. There are two factors that contribute to $\gamma_C$:
polarization of the 2DEG and polarization of the holes. Let us
consider an even electron filling factor when the electron gas is
composed of an equal number of electrons with the spin up and spin
down. Then, $\gamma_C$ is determined by the polarization of holes
only, since the electron gas is unpolarized. This is the case
considered in the present paper: the polarization of the
luminescence observed is caused by a thermal distribution of holes
on acceptor levels split by the magnetic field.

Tuning the 2DEG concentration can be achieved by the electrical
polarization of a semi-transparent electrode prepared on the sample
surface. We use this possibility to perform luminescence
measurements as a function of $n_s$ and $B$, and we show that
$\gamma_C$ depends on $n_s$ when $B$ and $T$ are constant.

\section{Experiment}

The sample under investigation was a high quality
GaAs/Al$_{0.33}$Ga$_{0.67}$As heterostructure grown on
semi-insulating GaAs substrate. The GaAs channel of about 1 $\mu$m
above a 50 periods of 5 nm/5 nm GaAs/AlAs superlattice contains
unintentional acceptors at a concentration less than 10$^{14}$
cm$^{-3}$. The AlGaAs barrier comprises an undoped 45 nm thick
AlGaAs spacer and a uniformly Si-doped 35 nm thick AlGaAs layer; the
doping level amounts to 10$^{18}$ cm$^{-3}$. The $\delta$-layer of
Be atoms with the concentration of 10$^9$ cm$^{-2}$ was introduced
into the GaAs channel at the distance $z_0$ = 30 nm away from the
GaAs/AlGaAs interface. The barrier was covered with a 15 nm thick
GaAs cap layer.

The measurements were carried out in an optical helium cryostat
supplied with a 5 T split-coil. All measurements were carried out at
1.6 K and the temperature was stabilized within 0.02 K by pumping
the helium gas through a manostat. The luminescence was excited by a
He-Ne laser. All data presented in this paper were obtained at the
same laser excitation power. Both cw and time-resolved measurements
were performed. The power of excitation was a few mW/cm$^{2}$ and
was a few orders of magnitude smaller than that corresponding to a
saturation of the luminescence signal.  In the case  of time -
resolved measurements, laser pulses were generated by passing the
laser beam through an acoustooptical modulator driven by a generator
of rectangular voltage pulses. The time resolution was 5 ns. The
luminescence passed through a $\lambda/4$ plate followed by a quartz
linear polarizer. The circular polarizations were separated by
turning the $\lambda/4$ plate. The luminescence was analyzed by a
spectrometer supplied with a CCD camera (for cw measurements) or
with a photomultiplier (for time-resolved measurements). In the
latter case, the photomultiplier signal was amplified and directed
to a photon counter.

A semi-transparent Au gate electrode and an ohmic contact were
fabricated on the sample surface and the concentration of the 2DEG
was tuned  by polarizing the gate in the backward direction. $n_s$
was estimated by determination
 the magnetic field, $B_{\nu = 2}$, at which the luminescence from the $N$=1 LLs
disappears (two Landau levels correspond to each $N$). Then
$n_s=2B_{\nu=2}/(h/e)$, where $e$ is the electron charge and $h$ is
the Planck constant.

\section{Results}

\begin{figure} \begin{center}
\includegraphics*[scale=0.8,angle=0]{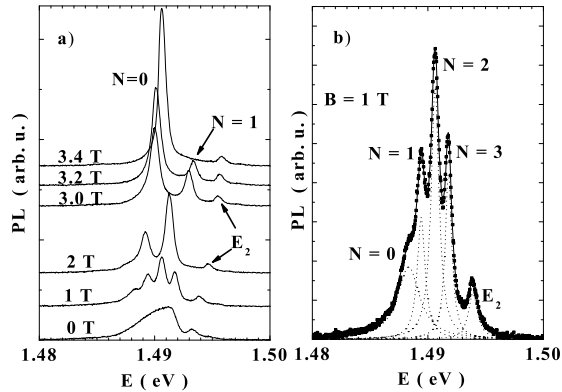}
\caption{a) Evolution of the luminescence spectrum with the magnetic
field $B$ (indicated). LLs of the ground electrical subband and the
peak corresponding to the first excited ($E_2$) subband are visible.
LLs are labeled with their number, $N$. In this case, the $N=1$ peak
disappears at $B_{\nu=2} \approx 3.3$ T. The spectra are vertically
shifted for better presentation. b) An example of deconvolution of a
luminescence spectrum into Lorentzians.} \label{JL2}
\end{center} \end{figure}

\begin{figure} \begin{center}
\includegraphics*[scale=0.8,angle=0]{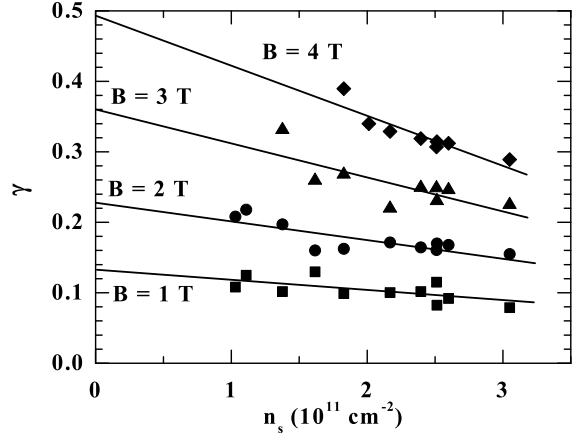}
\caption{$\gamma_C(n_s)$ dependence for 1 T (squares), 2 T
(circles), 3 T (triangles) and 4 T (diamonds). Solid lines are
linear extrapolations.  Estimated error of $\gamma_C$ is 0.03.}
\label{JL3}
\end{center} \end{figure}

An example of the magnetic field evolution of the luminescence
spectrum is shown in figure \ref{JL2}a. With an increase of $B$, the
number of populated LLs of the first electrical subband decreases as
their degeneracy grows.
The analysis of the polarization of the luminescence starts with a
deconvolution of each spectrum into separate Lorentzian peaks
corresponding to pairs of LLs (figure \ref{JL2}b). We subtract from
the total spectrum Lorentzians corresponding to the second
electrical subband and the highest in energy pair of LLs of the
first electrical subband ($E_2$ and $N=3$ peaks in figure
\ref{JL2}b). This leaves that part of the spectrum which corresponds
to an equal number of LLs occupied with spin-up and spin-down
electrons, i.e., to the totally unpolarized electron gas. The area
of that part of the spectrum is proportional to the intensity of the
luminescence, and used to calculate $\gamma_C$. This procedure
allows to determine $\gamma_C$ at given $B$ for different $n_s$. The
results are shown in  figure \ref{JL3}. Clearly, $\gamma_C$ depends
on $n_s$, and to interpret this result is the purpose of the present
paper.

\begin{figure} \begin{center}
\includegraphics*[scale=0.8,angle=0]{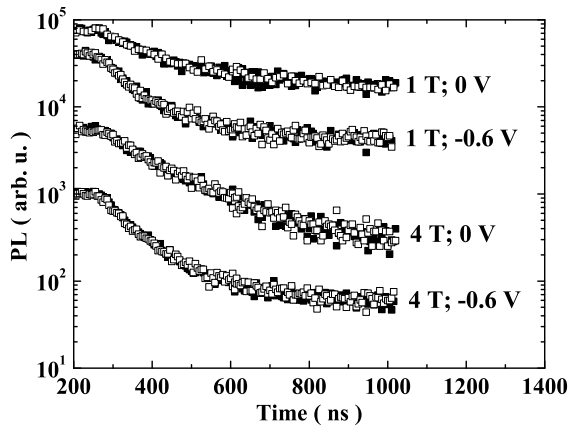}
\caption{Examples of a time dependence of the $\sigma^-$ (solid
squares) and $\sigma^+$ (open squares) components of the
luminescence from $N=0$ LLs for a few values of $B$  and applied
gate polarizations (indicated).  Each $\sigma^-$ signal was scaled
to coincide with the corresponding $\sigma^+$ signal. The exciting
laser pulse ends at about 270 ns.} \label{JL4}
\end{center} \end{figure}

Figure \ref{JL4} shows examples of a temporal evolution of the
luminescence signals measured in $\sigma^+$ (open squares) and
$\sigma^-$ (solid squares) polarizations. The $\sigma^-$ data was
multiplied by a constant factor to coincide with the $\sigma^+$
data. Corresponding $\sigma^+$ and $\sigma^-$ signals are
proportional in the whole time domain, both within the laser pulse,
and after the pulse end. The degree of polarization does not depend
on time (within the time resolution of the present experiment) which
indicates that the system is stationary and time-resolved
polarization measurements give the same value of $\gamma_C$ as cw
measurements. We used this fact to analyze polarization data
obtained from cw measurements which essentially improved the signal
to noise ratio.

The fact that the system investigated is stationary does not mean
that the distribution of holes on acceptor levels correspond to the
temperature of the helium bath surrounding the sample. We refer here
to time-resolved polarization studies on similar structures
\cite{Filin} which show that at 1.6 K the relaxation time of
photoexcited holes on the acceptor levels is of the order of
$10^{-10} - 10^{-9}$ s. This is a few orders of magnitude shorter
than the luminescence decay time which is of the order of $10^{-7}$
s, as can be concluded from figure \ref{JL4}. For this reason, we
can assume that photoexcited holes are distributed on the acceptor
levels according to an equilibrium thermal distribution
corresponding to the helium bath temperature of 1.6 K.

The problem of distribution of electrons is avoided in the present
considerations because we take into account the luminescence
originating from fully occupied Landau levels only.

\section{Stark effect on a bound $\Gamma_8$ hole}

To interpret $\gamma(n_s)$ dependence shown in figure \ref{JL3}, we
begin with an analysis of the Stark effect on the acceptor bound
hole. This problem was recently considered by Smit {\em et al.}
\cite{Smit} who calculated the energy  of acceptor levels in the
electric field. Our approach is complementary: we take into account
the magnetic field, not considered in  \cite{Smit}, and we discuss
an influence of mixing of hole levels by the electric field on
$\gamma_C$.

A group theory based Hamiltonian for the $\Gamma_8$ hole in the
electric ($F$) and magnetic field ($B$)was introduced by Bir,
Butikov and Pikus \cite{BirButikovPikus1, BirButikovPikus2}. In the
case of both $F$ and $B$ in the $z$ direction, one gets the
following perturbation Hamiltonian of the linear Zeeman and linear
Stark effects:
\begin{eqnarray}
H^{'} =   \mu_0(g_1^{'} J_z + g_2^{'}J_z^3)B +
\frac{p}{\sqrt{3}}F(J_xJ_y+J_yJ_x),
 \label{hamilton}
\end{eqnarray}
where  $\mu_0$ is the Bohr magneton, $g_1^{'}$ and $g_2^{'}$ are
isotropic and anisotropic $g$-factors, respectively, $J_i$ are the
$J$ = 3/2 orbital momentum matrices, and $p$ is the dipole moment of
Be acceptor.

Let us denote the unperturbed acceptor wavefunctions as $a\phi_i$,
$i=1, ..., 4$, where $a$ is an envelope function. We choose the
following  basis  of the wavefunctions $\phi_i$: $\phi_1=
\frac{1}{\sqrt{2}}(X+\mbox{i}Y)\alpha, \phi_2=
\frac{\mbox{i}}{\sqrt{6}}((X+\mbox{i}Y)\beta-2Z\alpha), \phi_3=
\frac{\mbox{i}}{\sqrt{6}}((X-\mbox{i}Y)\alpha+2Z\beta),\phi_4=
\frac{\mbox{i}}{\sqrt{2}}(X-\mbox{i}Y)\beta$ (corresponding to the
magnetic quantum number $m_J$= 3/2, 1/2, - 1/2 and -3/2,
respectively). $\alpha$ and $\beta$ are spin $\frac{1}{2}$ up and
down spinors, and $X$, $Y$ and $Z$ are functions transforming like
$x$, $y$ and $z$ under operations of the $T_d$ symmetry group,
respectively \cite{Bhattacharjee}. The above Hamiltonian takes the
form:
\begin{equation}
H^{'} = \left[ \begin{array}{rrrr}
b_1 & 0 & \mbox{i}\epsilon_L & 0 \\
0 & b_2 & 0 & \mbox{i}\epsilon_L \\
-\mbox{i}\epsilon_L & 0 & -b_2 & 0 \\
0 & -\mbox{i}\epsilon_L & 0 & -b_1
\end{array} \right]
\label{hamiltonmatrix}
\end{equation}
where $b_1 =\mu_0(\frac{3}{2}g_1^{'} + \frac{27}{8}g_2^{'})B$, $b_2
=\mu_0(\frac{1}{2}g_1^{'} + \frac{1}{8}g_2^{'})B$ describe the
linear Zeeman effect  and $\epsilon_L = p F$ describes the linear
Stark effect.

 The Hamiltonian given by equation
\ref{hamiltonmatrix} can be diagonalized analytically and explicit
formulae for mixed wave functions ($\Psi_i$) are:
$\Psi_{1,3}=Ca(\phi_{1,3}+\rho \phi_{3,1})$ and
$\Psi_{2,4}=Ca(\phi_{2,4}+\rho \phi_{4,2})$, where $C$ is a
normalizing constant and $\rho$ describes mixing of the wave
functions; $\rho = \mbox{i}\alpha(1-\sqrt{1+\alpha^{-2}} )$, where
$\alpha = \mu_0(g_1^{'}+\frac{7}{4}g_2^{'})B/\epsilon_L$.
 The
corresponding energies are: $E_{1,3} = (b_1 -b_2 \pm
\sqrt{(b_1+b_2)^2+4\epsilon_L^2})/2$ and $E_{2,4} =(-b_1 +b_2 \pm
\sqrt{(b_1+b_2)^2+4\epsilon_L^2})/2$. The conduction band $\Gamma_6$
states are not affected by the electric field \cite{Smit}.

\begin{figure} \begin{center}
\includegraphics*[scale=0.8,angle=0]{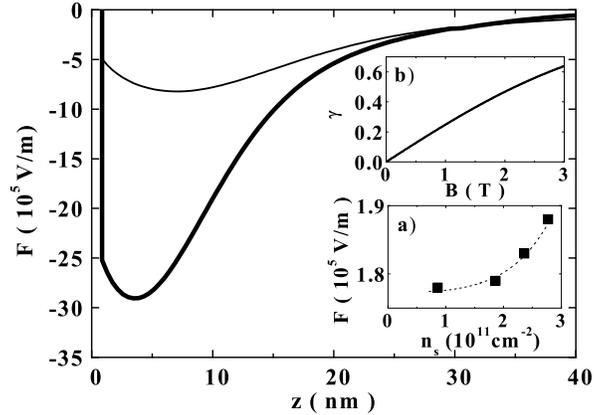}
\caption{The heterostructure electric field for $n_s= 0.35 \times
10^{11}$ cm$^{-2}$ (thin line) and $2.77\times 10^{11}$ cm$^{-2}$
(thick line). a) The electric field at $z_0$ = 30 nm as a function
of the electron concentration $n_s$. Dashed line is guide for eye.
b) $\gamma_C(B)$ dependence calculated on the basis of energies and
wave functions obtained by diagonalization of the Hamiltonian given
by Eq. \ref{hamiltonmatrix} at 1.6 K and for $p=1$ D. Data for $F$=
0 and $F = 1.9\times10^5$ V/m cannot be distinguished in the scale
of the figure. } \label{JL5}
\end{center} \end{figure}

In the presence of the electric field, the selection rules shown in
figure \ref{JL1}a, valid for $F = 0$,  are replaced by the scheme
shown in figure \ref{JL1}b which is a result of mixing of $m_J =
3/2$ with $m_J = -1/2$ and $m_J = 1/2$ with $m_J = -3/2$ states
described by the above Hamiltonian. In consequence, a decrease of
$\gamma_C$ in comparison with the $F=0$ case can be expected. To
determine the magnitude of this effect we use the functions $\Psi_i$
to calculate the intensity of the $\sigma^+$ and $\sigma^-$
transitions in the following way.

According to the time-dependent perturbation theory, the intensity
of an optical transition between the states $|\psi_1\rangle$ and
$|\psi_2\rangle$ is proportional to
$|\langle\psi_1|\textbf{e}\cdot\textbf{p}|\psi_2\rangle|^2$, where
$\textbf{e}$ is the polarization vector of the electromagnetic wave
and $\textbf{p}$ is the electron momentum. In the case of a
transition between electrons occupying Landau levels and holes
occupying acceptor levels, the intensity observed in the
polarization $\textbf{e}$ is equal to
\begin{equation}
I_{\textbf{e}}= \Xi \sum_{\sigma}\sum_{i=1}^{4}w_\sigma
w_{i}|\langle\mu\sigma|\textbf{e}\cdot\textbf{p}|\Psi_i\rangle|^2
 . \label{intensity}
\end{equation}
$\Xi$ is a polarization independent factor, $\mu\sigma$ is the wave
function of a conduction band electron with the spin projection
$\sigma$ equal to $\alpha$ or $\beta$. Statistical weights $w_i$
describe the probability of occupation of each of four acceptor
Zeeman $i$: $w_i=\exp[(E_1-E_i)/k_BT]/\sum_i\exp[(E_1-E_i)/k_BT]$,
where $E_1$ is the energy of the lowest lying level, corresponding
to $m_J=3/2$. Statistical weights for electron spin levels,
$w_\sigma$, are equal to 1 because we consider equal number of fully
occupied Landau levels with spin up and down.

 Let
$I_{\sigma^{-}}=I_{(\textbf{e}_x-\mbox{i}\textbf{e}_y)/\sqrt{2}}$
and
$I_{\sigma^{+}}=I_{(\textbf{e}_x+\mbox{i}\textbf{e}_y)/\sqrt{2}}$.
Then
\begin{equation}
 \gamma_C= \frac{(I_{\sigma^-}-I_{\sigma^+})}
{(I_{\sigma^-}+I_{\sigma^+})}=
 \frac{-2\mbox{Im}\sum_{\sigma
i}w_{i}\langle\mu\sigma|p_x|\Psi_i\rangle\langle
\Psi_i|p_y|\mu\sigma\rangle}{\sum_{\sigma
i}w_{i}(|\langle\mu\sigma|p_x|\Psi_i\rangle|^2 + |\langle
\mu\sigma|p_y|\Psi_i\rangle|^2)}. \label{gammaC}
\end{equation}

In considering the Stark effect we assume in this Section that both
electron wavefunctions $\mu_0$ and hole wavefunctions $aX, aY$
transform according to representations of the $T_d$ symmetry group.
Matrix elements in Eq. \ref{gammaC} can be expressed by $A_{\mu_0}
\equiv \langle\mu_0|p_x|aX\rangle=\langle\mu_0|p_y|aY\rangle$. This
equality results from the assumed symmetry of wavefunctions and the
fact that the $x$ and $y$ directions are equivalent in the $T_d$
symmetry.   We get
\begin{eqnarray}
 I_{\sigma^-}=\Xi|C|^2 |A_{\mu_0}|^2
\left((w_1+|\rho|^2w_3) +\frac{1}{3}(w_2+|\rho|^2w_4)\right), \nonumber \\
\ \\
I_{\sigma^+}=\Xi|C|^2|A_{\mu_0}|^2\left(\frac{1}{3}(w_3+|\rho|^2w_1)
 +(w_4+|\rho|^2w_2)\right), \nonumber
\label{sigmapm}
\end{eqnarray}
where $|C|^2=(1+|\rho|^2)^{-1}$.

To calculate $\gamma_C$ as a function of $F$ using (\ref{sigmapm})
we estimate the heterostructure electric field in the plane where Be
acceptors are placed, $F(z_0)$. The distribution of charges and the
electrostatic potential in the heterostructure were calculated by a
self-consistent solution of the Schr\"{o}dinger and Poisson
equations. The overall charge neutrality was guarantied by taking
into account interface and surface charges, which simulate also the
effect of a gate electrode and which are used to control the
concentration of the 2DEG in the heterostructure. The results of the
calculations are the subband energies and wavefunctions as a
function of the total 2DEG concentration $n_s$. The calculations
show that at $z_{0}$ = 30 nm, where Be acceptors are placed, the
electric field is almost insensitive to $n_s$: it increases from
$1.8\times 10^5$ V/m to $1.9\times 10^5$ V/m when $n_s$ increases
from $10^{11}$ cm$^{-2}$ to $3 \times 10^{11}$ cm$^{-2}$ (see figure
\ref{JL5}).

 The $g$-factors of holes bound to Be acceptors in bulk GaAs were
  obtained in a far infrared experiment by Lewis,
Wang and Henini \cite{Lewis}: $g_1^{'} = 0.3$ and $g_2^{'} = 0.09$.
To calculate $\rho$, the value of the Be acceptor dipole moment,
$p$, should be known. Since, apparently, this data is not available
in the literature, let us assume that  $p$ = 1 D (1 D = 3.3 $\cdot$
10$^{-30}$ Cm). This value is higher than the dipole moment of
several different acceptors in bulk Si investigated by K\"{o}pf and
Lassman \cite{KopfLassmann}, which was found to fall in the range
0.26 D - 0.9 D. Taking $p = 1$ D, $T = 1.6$ K and $F = 1.9\times
10^5$ V/m we can estimate the upper limit of the Stark effect energy
corresponding to the present experimental conditions. We get
$\epsilon_L \sim$ 5 $\mu$eV, which is negligible in comparison with
the thermal energy at 1.6 K. It means that the distribution of holes
over the Zeeman levels is not changed by  the heterostructure
electric field. An influence of mixing of hole levels by the
electric field on $\gamma_C$ can be investigated by calculation of
$\gamma_C(B)$ for different values of $F$. Results for $F = 0$ and
$F=1.9 \times 10^5$ V/m are shown in the inset to figure \ref{JL5}.
The two curves practically coincide: for the magnetic field of
interest, $\gamma_C$ for these two values of $F$ differs by less
than $10^{-3}$.

An analysis presented in this Section shows that it is not possible
to explain the $\gamma_C(n_s)$ dependence by the Stark effect:
neither energy shifts of hole levels nor mixing of levels is strong
enough to influence $\gamma_C$ significantly.

\section{Circular polarization of luminescence from systems of the $C_{2v}$ symmetry}

Let us analyze  now the influence of the low symmetry of the
heterostructure potential on $\gamma_C$. In structures of the
$C_{2v}$ symmetry, the directions $x'$ ([1$\overline{1}$0]) and $y'$
([110]) are  not equivalent which leads to an anisotropy of optical
transitions in the (001) plane. This leads to experimentally
observed nonzero degree of the linear polarization: $\gamma_L =
(I_{[110]} - I_{[1\overline{1}0]})/(I_{[110]} +
I_{[1\overline{1}0]}) \neq 0$. It will be shown below that in
systems of the $C_{2v}$ symmetry, $\gamma_C$ and $\gamma_L$ are
related to each other:$(\gamma_C/\gamma_{C0})^2+ \gamma_L^2 = 1$,
where $\gamma_{C0}$ is the degree of circular polarization for the
$T_d$ symmetry. The increase of  the strength of  perturbation
causing $T_d \rightarrow C_{2v}$ symmetry lowering results in an
increase of $\gamma_L$ and a corresponding decrease of $\gamma_C$.
For this reason, effects of the symmetry lowering can be
investigated by an analysis of $\gamma_C$.

A basis for modeling the relation between a low symmetry of a
quantum system and its optical anisotropy is a theory of Ivchenko
{\em et al.} \cite{Ivchenko2}. According to this theory, the
anisotropy originates from mixing of light and heavy valence band
holes while it is  explicitly assumed that the $T_d$ symmetry of
conduction band electron wavefunctions is preserved. R\"{o}sller and
Kainz, starting from the theory of Ivchenko {\em et al.}, showed
that the $\mathbf{kp}$ coupling of the valence and conduction bands
leads to [110] vs. [1$\overline{1}$0] anisotropy of the conduction
band wavefunctions \cite{Rossler}. These calculations were carried
out for the $B=0$ case only.

A special feature makes the heterostructure investigated in this
work different from other systems of the $C_{2v}$ symmetry studied
by other groups. The point is that holes participating in  optical
transitions  are localized as far as 30 nm from the interface (i.e.,
about 10 Bohr radii) and the influence of the localized interface
potential on the holes can be neglected. We showed that the
long-range heterostructure potential is also too weak to mix hole
levels significantly. Other mixing mechanisms, like the internal
stress or stress-related piezoelectric effects, even if present, do
not depend on the electron concentration.

For this reason, we propose that the anisotropy results from a low
symmetry of conduction band electron wavefunctions. We propose that
the source of the anisotropy is a perturbation of the valence band
induced both by the localized interface potential and the long-range
heterostructure electric field. These perturbations do not influence
holes localized on Be acceptors, as we argued above, but only the
valence band states in the region of the GaAs channel where the
heterostructure electric field is strong and the density of
conduction band electron wavefunction is large.  Next, the symmetry
of conduction band electron wavefunctions is lowered due to the
$\mathbf{kp}$ coupling of bands described in \cite{Rossler}. The
$\mathbf{kp}$ perturbation increases with the electron wavevector
which means that it becomes stronger when $n_s$ increases. We
propose that the $\gamma_C(n_s)$ dependence observed results from
changes of the heterostructure electric field correlated to changes
in the 2DEG concentration.

As in the theory of \cite{Ivchenko2}, we point at two factors
contributing to the symmetry lowering: the long-range
heterostructure electric field (leading to the Pockels effect) and
the localized interface potential. In fact, figure \ref{JL5} shows
that for $z<20$ nm a large increase, of the order of 10$^6$ V/m, of
the heterostructure electric field is observed with the increase of
$n_s$. This means that a perturbation of the valence band, and in
consequence, of the conduction band, increases with $n_s$. Also,
when $n_s$ increases, the heterostructure well becomes narrower,
electron wavefunctions shift towards the interface and the influence
of the localized interface potential on the electron states
increases. The latter mechanism should lead to much stronger effects
in the case of the single interface in the heterostructure
investigated in comparison to quantum wells. The reason is that in
quantum wells of the $C_{2v}$ symmetry, contributions of the two
interfaces enter with opposite signs and partially cancel each
other. Also, the fact that Be-bound holes are unperturbed, makes
effects of lowering the symmetry of the conduction band states more
evident.

Carrying out calculations similar to these presented in
\cite{Rossler} with taking into account the magnetic field is beyond
the scope of the present paper. We propose, however, a
phenomenological description of the polarization in the system
investigated which allows us to derive selection rules for optical
transitions between an electron of the $C_{2v}$ symmetry and a
bulk-like hole of the $T_d$ symmetry.

>From Eq. \ref{intensity} one can calculate the intensity of the
$\Gamma_6 \rightarrow \Gamma_8$ transition for the [110]
($\textbf{e} = (\textbf{e}_x + \textbf{e}_y)/\sqrt{2}$) and
[1$\overline{1}$0] ($\textbf{e}=(\textbf{e}_x -
\textbf{e}_y)/\sqrt{2}$) polarizations to get
\begin{equation}
I_{[110]} - I_{[1\overline{1}0]} = 2\Xi\mbox{Re}\sum_{\sigma
i}w_{i}\langle\mu\sigma|p_x|a\phi_i\rangle\langle
a\phi_i|p_y|\mu\sigma\rangle. \label{realis}
\end{equation}
In the $T_d$ symmetry, the directions [110] and [1$\overline{1}$0]
are equivalent ($\gamma_L$ = 0 in this case) and the above
expression is equal to zero. This is possible only when
\begin{equation}
\begin{array}{l}
\langle\mu_0|p_x|aX\rangle=\langle\mu_0|p_y|aY\rangle \\
\langle\mu_0|p_x|aY\rangle=0 \\
\langle\mu_0|p_y|aX\rangle=0.
\end{array}
\label{matrixTd}
\end{equation}
 To get a non zero
value of $\gamma_L$ with perturbed electron wavefunctions $\mu$ of
the $C_{2v}$ symmetry, it is necessary that
\begin{equation}
\begin{array}{l}
\langle\mu|p_x|aX\rangle=\langle\mu|p_y|aY \rangle\\
\langle\mu|p_x|aY\rangle\neq 0 \\
\langle\mu|p_y|aX\rangle\neq 0,
\end{array}
\label{inequality}
\end{equation}
where the first equation reflects the equivalence of the [100] and
[010] directions in the $C_{2v}$ symmetry.

Calculation of matrix elements
$\langle\mu\sigma|p_x\pm\mbox{i}p_y|a\phi_i\rangle$ with taking into
account  (\ref{inequality}), gives the selection rules of optical
transitions observed in the circular polarizations in the $C_{2v}$
symmetry. Non zero intensities are shown by arrows in figure
\ref{JL1}b. Lowering the symmetry of conduction band wavefunctions
leads to relaxation of selection rules and a decrease of $\gamma_C$
caused by mixing of $\sigma^-$ and $\sigma^+$ transitions.

To quantify the model, we introduce a real (see below) parameter
$\zeta$ and we assume that
\begin{equation}
\begin{array}{l}
\langle\mu|p_x|aX\rangle=\langle\mu|p_y|aY\rangle=A_\mu, \\
\langle\mu|p_x|aY\rangle=\langle\mu|p_y|aX\rangle = \zeta A_\mu.
\end{array}
\label{model}
\end{equation}
$\zeta$ describes an evolution of matrix elements when the symmetry
is lowered. We get
\begin{equation}
\begin{array}{l}
I_{\sigma^-}=(1+\zeta^2)(R+S)+ (1-\zeta^2)(R-S) \\
I_{\sigma^+}=(1+\zeta^2)(R+S)- (1-\zeta^2)(R-S),
\end{array}
\label{IplusIminus}
\end{equation}
where
\begin{equation}
\begin{array}{l}
R=\Xi|A_\mu|^2(\frac{1}{2}w_1+\frac{1}{6}w_2) \\
S=\Xi|A_\mu|^2(\frac{1}{6}w_3+\frac{1}{2}w_4).
\end{array}
\label{RS}
\end{equation}
The degree of circular polarization is
\begin{equation}
\gamma_C=\frac{1-\zeta^2}{1+\zeta^2}\frac{R-S}{R+S},
 \label{gammaCcalc}
\end{equation}
while $\gamma_L = 2\zeta/(1+\zeta^2)$. (Assuming that $\zeta$ is
complex, one can calculate the degree of linear polarization between
the [100] and [010] directions:
$(I_{[100]}-I_{[010]})/(I_{[100]}+I_{[010]})=-\frac{2
\mbox{\tiny{Im}} \zeta}{1+\zeta^2}\frac{R-S}{R+S}$ which is equal to
zero because these directions are equivalent. This implies
$\mbox{Im}\zeta=0$.)

Assuming that $A_\mu=(1+\zeta)A_{\mu_0}$, where $\mu_0$ transforms
into $\mu$ when the symmetry lowers from $T_d$ to $C_{2v}$,
$(R-S)/(R+S)$ does not depend on $\zeta$ and is equal to
$\gamma_{C0}$ - the degree of circular polarization that would be
measured in the case when both the electron and hole states were of
the $T_d$ symmetry. These allow us to derive the relation between
$\gamma_L$ and $\gamma_C$ mentioned above.

\begin{figure} \begin{center}
\includegraphics*[scale=0.8,angle=0]{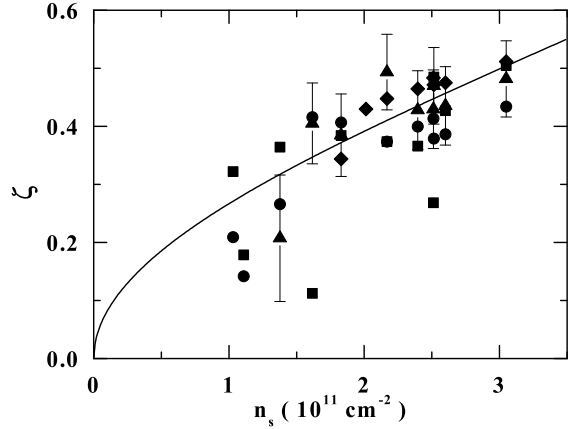}
\caption{$\zeta(n_s)$ calculated from data in figure \ref{JL3}.
Squares: 1 T, circles: 2 T, triangles: 3 T, diamonds: 4 T. As an
example, solid line correspond to the linear fit for $B$ = 3 T in
figure \ref{JL3}. For clarity, error bars for $B$ = 3 T data only
are shown. The error of $\zeta$ corresponds to the error of
$\gamma_C$ equal to 0.03.} \label{JL6}
\end{center} \end{figure}

To estimate the range of $\zeta$ corresponding to $\gamma_C$
measured, we use Eq. \ref{gammaCcalc} which gives
$\zeta=\sqrt{\frac{1-\kappa}{1+\kappa}}$, where
$\kappa=\gamma_C/\gamma_{C0}$. With data of figure \ref{JL3} and
$\gamma_{C0}=\gamma_C(n_s=0)$, we get results presented in figure
\ref{JL6} showing $\zeta$ increasing from about 0.2 to 0.5 with
$n_s$ increasing from 10$^{11}$ cm$^{-2}$ to $3\times 10^{11}$
cm$^{-2}$.

The results presented in this Section allow us to suggest that a
widely used approximation of conduction band electron wavefunctions
as unperturbed functions of the $T_d$ symmetry is not generally
valid. The symmetry lowering of conduction band states results from
the perturbation of the valence band by a long-range heterostructure
electric field and the localized interface potential. Measurements
on heterostructures with different barrier height could help to
separate these two contributions. This is because the role of the
localized interface potential should strongly increase with lowering
the barrier height due to an exponential increase of penetration of
the barrier by electron wavefunctions.

\section{Summary and conclusions}

Low temperature polarized magnetoluminescence experiments on a high
quality GaAs/AlGaAs heterostructure with Be acceptor $\delta$-layer
incorporated into the GaAs channel were carried out. A metallic gate
on the sample surface allowed to tune the 2DEG concentration in the
heterostructure. The degree of circular polarization of the
luminescence originating from transitions between 2DEG and
photoexcited holes bound to Be acceptors was analyzed. We observed
that $\gamma_C$ decreases with the increase of $n_s$. In principle,
this effect can be explained by the Stark effect on Be - localized
hole and/or by an in-plane anisotropy of optical transitions,
reflecting the $C_{2v}$ symmetry of the heterstructure potential. An
analysis of the Stark effect was based on a group theory
Hamiltonian, with taking into account results of self-consistent
calculations of the heterostructure electric field. It was shown
that the Stark effect can be neglected in the heterstructure
investigated, which led to the conclusion that the effect
responsible for the $\gamma_C(n_s)$ dependence is a low symmetry of
the heterostructure potential. We argued that Be-localized holes are
of the $T_d$ symmetry which led us to the conclusion that the
optical anisotropy is related to a low symmetry of the conduction
band electron wavefunctions. We propose that this symmetry lowering
results from a perturbation of the valence band which induces
perturbation of the conduction band due to the $\mathbf{kp}$
coupling. We used symmetry arguments to derive selection rules of
optical transitions in the case when the conduction band electrons
and holes are of the $C_{2v}$ and $T_d$ symmetry, respectively, and
we provided a phenomenological description of the experimental data.

In conclusion, it is proposed that due to a large separation of
holes from the interface, the in-plane anisotropy observed is due to
lowering of the symmetry of the conduction band electron
wavefunctions only. The mechanism responsible for the symmetry
lowering is a perturbation of the valence band states which
influences the conduction band states through the $\mathbf{kp}$
coupling. The increase of the perturbation with $n_s$ is caused by
an increase of the heterostructure electric field with $n_s$.
Measurements on heterostructures with different barrier heights
could help to separate a contribution from the localized interface
potential from that of the long-range electric field. 
\begin{acknowledgments}
 J. {\L}. is thankful to M.~ Grynberg and A.~ {\L}usakowski for discussions.
\end{acknowledgments}

\end{document}